\documentclass{article}

\usepackage{arxiv}

\usepackage[utf8]{inputenc} % allow utf-8 input
\usepackage[T1]{fontenc}    % use 8-bit T1 fonts
\usepackage{hyperref}       % hyperlinks
\usepackage{url}            % simple URL typesetting
\usepackage{booktabs}       % professional-quality tables
\usepackage{amsfonts}       % blackboard math symbols
\usepackage{nicefrac}       % compact symbols for 1/2, etc.
\usepackage{microtype}      % microtypography
\usepackage{lipsum}
\usepackage{graphicx}
\graphicspath{ {./images/} }
\usepackage{amsmath}

\title{Simulators for Quantum Network Modelling: A Comprehensive Review}

\author{
 Oceane Bel \\
  Physical \& Computational Sciences Directorate\\
  Pacific Northwest National Laboratory\\
Richland, WA, USA \\
  \texttt{obel@pnnl.gov} \\
   \And
 Mariam Kiran \\
  Computational Sciences and Engineering Division\\
 Oak Ridge National Laboratory\\
 Oak Ridge, TN, USA \\
  \texttt{kiranm@ornl.gov} \\
}

\begin{document}
\maketitle
\begin{abstract}
Quantum network research, is exploring new networking protocols, physics-based hardware and novel experiments to demonstrate how quantum distribution will work over large distances. Current work explores much of these concepts in simulations, that are developed to understand how quantum networking will be set up and researchers can experiment virtually. Exposing flaws in network designs, like unsustainable topologies, or develop protocols that efficiently utilize network resources, simulators can also help assess whether workloads are balanced across virtual machines in the network. However, much of these simulation models come without reliable verification methods, for testing performance in real deployments.

In this paper, we present a review of, to the best of our knowledge, currently used toolkits for modeling quantum networks. With these toolkits and standardized validation techniques, we can lay down the foundations for more accurate and reliable quantum network simulators.

\end{abstract}

% keywords can be removed
%\keywords{First keyword \and Second keyword \and More}

\textbf{Topics: Quantum network, network models, simulators, verification and validation
}

%\section{Introduction}

\section{Introduction}
%\MK{Rewrite the starting of the intro.
%\\
%What is Quantum networking and why is it important? add a reference to recent paper about it
Quantum networks are the next generation of communication networks that deliver entanglement and connect distributed quantum physics devices such as quantum computers, sensors and detectors ~\cite{chiribella2009theoretical,wei2022towards,kozlowski2019towards, Peters:2023mpv}. Instead of classical `bits' (0s,1s), qubits are transmitted that can encode more information, by using the spin of the photons. Argued to be the next internet revolution, distributed quantum networks have shown case studies to process and communicate secure and highly accurate measurements, currently beyond the classical machines \cite{doi:10.1021/acsphotonics.9b00250}.

Industry quantum efforts are developing new hardware, protocols and tools that can enable quantum information exchange. These technologies can demonstrate reliable quantum communication with high rates of fidelity and automated error correction ~\cite{braunstein1998quantum,sidhu2021advances,muralidharan2016optimal}. Some of these approaches use discrete and continuous variable demonstrations, each bringing their own capabilities in the experiments \cite{Alshowkan:2021kpt}, or transmission over fiber optical cable or free space point to point communication. Network loss plays a huge role in guaranteeing the validity of the quantum states, current demonstrations have only shown upto 300 km \cite{Rao:2023exv}. With the help of quantum repeaters, one can extend these to longer distances such that quantum states can be refreshed or preserved \cite{azuma2023quantum,singh2021quantum,desef2021protecting}.

Quantum network simulations are also being developed to help define use cases, and collect data to build reliable devices. Several efforts such as Qunet~\cite{diadamo2021qunetsim}, ComNetsEmu~\cite{fitzek2020computing}, or Cisco's QnetLab are focused on developing kits that can interface with other simulations and provide a GUI, to build topology, collect parameters and investigate new protocols. These can demonstrate large-distance transmission using repeaters and help identify solutions to protocols that can utilize network resources effectively. However, these simulation toolkits, still lack robust testing and verification methods to verify the accuracy of simulated quantum networks and how these would translate in the real world hardware. 

\begin{figure*}[!htb] 
  \centering
  \includegraphics[width=1\linewidth]{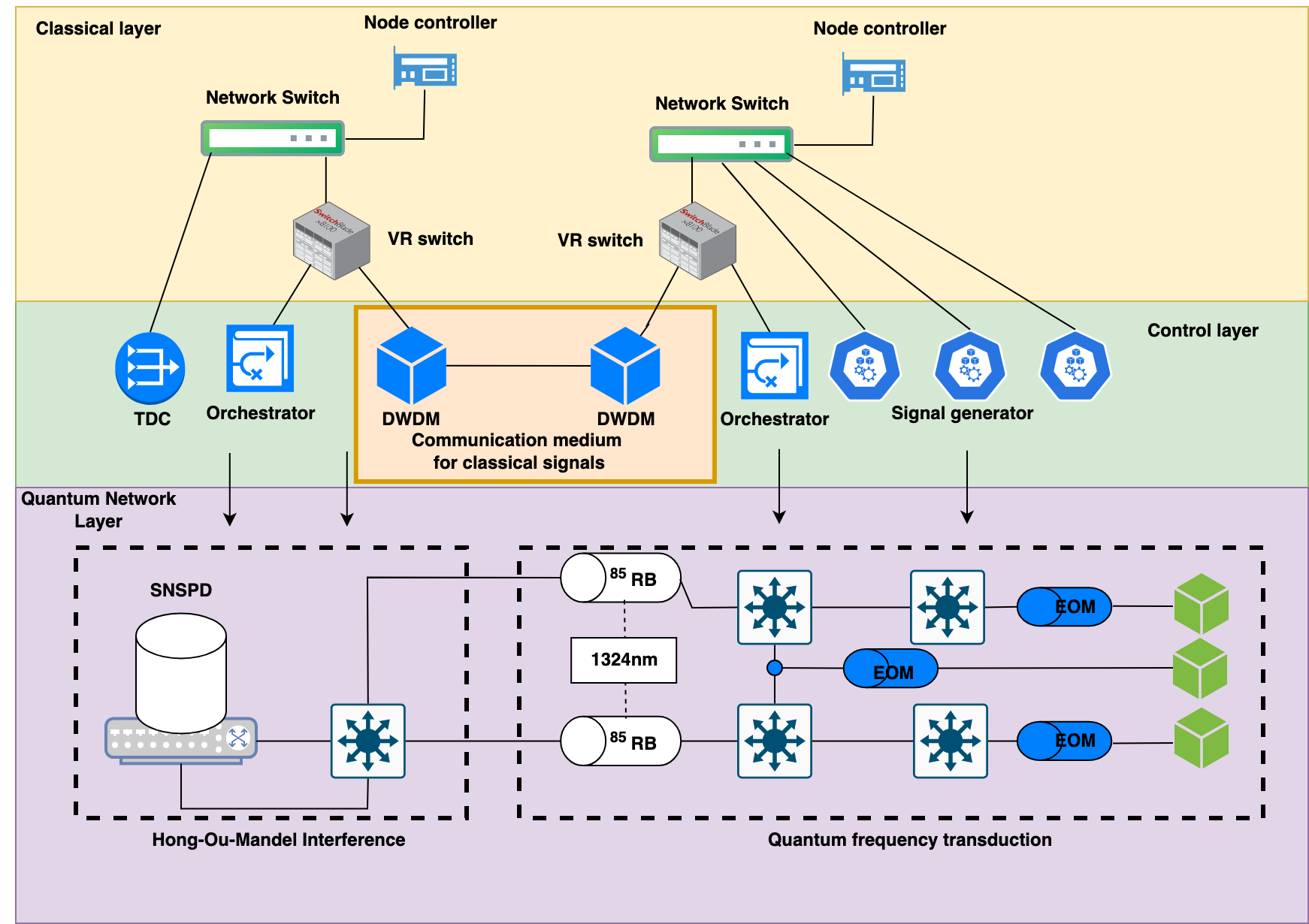}
  \caption{An example of interfacing quantum over optical network layer. This includes: The DWDM (Dense Wavelength Division Multiplexing) exists in the physical layer of classical optical networks. TDC (Time-to-Digital Converter) is mainly used for synchronization between different nodes. EOM (Electro-Optic Modulator) manipulates the light signals to convert the frequency of a photon while preserving its quantum information. SNSPD (Superconducting Nanowire Single-Photon Detector) help perform measurements, Hong-Ou-Mandel (HOM) interference and RB (Rubidium) is an alkali metal element.}
  \label{fig:NetworkLayer}
\end{figure*}

Developing quantum networks presents significant challenges for network providers and researchers.
Some exploration in monitoring equipment and tools~\cite{amoretti2019enhancing,rao2018control,khalid2023quantum} are integrating quantum and optical channels, often with a transduction layer for quantum network overlay (Figure~\ref{fig:NetworkLayer}). Here, transduction is applied to quantum information transmitted from one physical system to another, where qubit frequencies are converted to optic and vice versa, utilizing the underlying infrastructure.

%What does QN offer?
%Quantum networks (QN) use entanglement to encrypt data in a way that is mathematically proven unbreakable, revolutionizing everything from financial transactions to national security. By attempting to eavesdrop on the data being transferred from one point to another in the network the attacker risks destroying it even before they can get any information out of the data. Additionally, quantum channels offer the potential for faster, more efficient data transmission with reduced noise and interference. This could boost communication speeds and open up new possibilities for real-time collaboration.
%\\However where are the issues that are still being developed?

% Add the key words in the caption

%\MK{Add diagram of Quantum network on optical layer}

% Then, a simulator can help to the developer for finding the weaknesses that make a solution ineffective such as:
% •
% Is it based on suitable topology?

% •
% Is it utilizing the network resource effectively?

% •
% Is the workload distributed well between VMs?
In this paper, we investigate and compare quantum network simulators that are being used to develop key concepts in quantum networking. These simulators model new technologies, use analytical models and compare physics operations derived from established equations. Having a standardized validation method will allow consistency within the research community. This survey focuses on reviewing existing theoretical works for simulating qubit behavior on communication channels, discussing on what is further needed in this space.

\subsection{Motivation and Contributions}
% Establish Use cases to test simulators
% Develop benchmarks for simulations

The goal of this paper is to review the current quantum network simulation toolkits and understand how these are being used. Identifying the use cases being modelled, the paper aims to identify the limitations each toolkit has. This paper serves as an introduction to the quantum network simulation research, covering the current landscape and identifying future needs for validating quantum network simulators. 

In terms of validation, the paper aims to explore simple analytical models on how qubit models can be generated and tested for quantum networking. The paper is organized as follows: Section 2 discusses what quantum networks are designed for and its applications. Section 3 goes into details of example simulation platforms and their developments. Section 4 discusses the challenges in these platforms and how these can improved in Section 5. Finally, Section 6 concludes and presents future research directions in this area.

\section{Challenges in Designing Quantum Network Simulators}

The first design consideration is correctly modeling Quantum Nodes and Channels~\cite{dasari2016programmable}. Factors such as qubit connectivity, gate fidelity, and quantum channel characteristics~\cite{matsuo2019simulation}, and scaling with the size of the quantum network while maintaining computational efficiency are all challenges for simulators.

The second consideration is realism~\cite{fortes2015fighting} of the behavior of quantum operations on the network. Realistic quantum operations, such as gates, measurements, and entanglement generation, need to pay attention to the impact of noise and errors. This means that simulators need to have capabilities for simulating quantum error correction techniques. Using accurate quantum models to reflect the behavior of physical quantum systems, needs a tool to evaluate the performance of error correction codes and their effectiveness in mitigating the impact of these on quantum communication.

A third consideration, there is a need to develop a comprehensive benchmarking and validation framework to assess the performance of the simulator. This framework should include metrics for success rates, fidelity, speed, scalability, and resource utilization in entanglement distribution and other quantum communication tasks. It should also have a user-friendly interface, which allows users to setup, configure, and analyse quantum network simulations. It should provide tools for visualizing network states, quantum operations, and simulation results. The interface should be able to measure the accuracy of the simulation compared to what is expected in the real world. Existing work~\cite{vcepaite2022simulation} on validation of simulation has used simulator platforms, such as the QISKit platform developed by IBM. The interface needs to validate the simulator against experimental data where possible. Because of a lack of available hardware, gathering heuristics to create a theoretical representation of the network's expected behavior will help validating results. 

Finally, since work on quantum networks is continuously updating, simulators need to use a modular and extensible architecture. This allows users to easily incorporate new features, quantum algorithms, or simulation models to adapt the simulator to evolving research needs. Additionally, it also allows for validation tools to be incorporated into the simulation. Therefore, any standardized validation framework should also be modular to allow for new validation techniques to be added to the framework as they become available.

%          Title Designing a quantum network protocol
%            URL https://dl.acm.org/doi/abs/10.1145/3386367.3431293
%           Year 2020
%      Citations 93
%       Versions 9
%     Cluster ID 283984029953669139
%       PDF link https://dl.acm.org/doi/pdf/10.1145/3386367.3431293
% Citations list https://scholar.google.com/scholar?cites=283984029953669139&as_sdt=2005&sciodt=0,5&hl=en
%  Versions list https://scholar.google.com/scholar?cluster=283984029953669139&hl=en&as_sdt=0,5
%        Excerpt … (3) We evaluate the effectiveness of the proposed protocol against decoherence in a quantum network simulator. (4) We show that it remains functional on extremely limited near-term …

\subsection{Understanding the output of a simulation}
Simulations can be compared with known analytical solutions for specific quantum algorithms or network scenarios. For modeling entanglement distribution, validation methods include comparing the success rates, fidelity, and speed of entanglement generation with theoretical expectations and experimental data.

Another approach to validation is compare the resulting behavior of the simulation to an ideal solution. For this case, idealized simulations, where noise and errors are minimized or absent, can be used to verify the correctness of the implementation. Comparing results from a noise-free simulation with theoretical expectations helps ensure the accuracy of the simulator's fundamental quantum operations. Simulators like QuNetSim or SeQUeNCE, have assumed minimal to no errors in their models. 

\subsection{Quantum-Classical Hybrid simulators}
Researchers have combined classical and quantum hybrid networks to develop full quantum networks~\cite{diadamo2021integrating}. Here a validation framework also validates classical network capabilities and their impact on quantum networks. The interaction between classical and quantum networks includes simulating the interaction between classical and quantum information processing. Many existing quantum workloads and applications use a mix of pure quantum algorithms and classical programs. For example, ``proper quantum'' algorithms were still hybrid~\cite{weder2021hybrid} use the Shor algorithm to model quantum parts but used classical processing for the data.

Quantum machine learning~\cite{biamonte2017quantum,zhang2020recent} uses classical machine learning enhanced or replaced by quantum algorithms~\cite{ciliberto2018quantum,schuld2019quantum}. Hybrid approaches often involve classical pre-processing or post-processing steps alongside quantum algorithms to address machine learning tasks such as classification, clustering, and optimization. The preprocessing and optimization may take advantage of the cloud capabilities as an intermediate to distribute such large models developing Quantum Cloud Computing.

For Quantum Cloud Computing~\cite{singh2014quantum} to have the needed resources, quantum processors can be integrated into cloud computing environments. Researchers have explored how classical and quantum resources can be orchestrated to perform computations efficiently, taking advantage of the strengths of both. Companies such as IBM~\cite{castelvecchi2017ibm}, Microsoft~\cite{hooyberghs2022azure}, and Google~\cite{kaiiali2019cloud} have started offering cloud services that include access to quantum processors. These services allow users to run quantum algorithms and experiments on real quantum hardware through cloud platforms~\cite{luong2023towards}. The concept of hybrid quantum-classical cloud computing involves combining classical computing resources with quantum processing units. This approach is particularly useful for solving complex problems that leverage both classical and quantum algorithms.

\section{Overview of Quantum Networking}
%What is it
%What is it trying to acheive.
Quantum networking is the infrastructure connecting one or more quantum physics devices such as for distributed quantum computing, sensors or setting up an entangled quantum network. As a means to advance science, quantum networking applications are exploring how to connect multiple observatories for super-resolution images of distant planets~\cite{khabiboulline2019optical,sidhu2021advances}, or even link high-powered microscopes for unprecedented views of the micro-organisms~\cite{yu2023telecom}, to help find more key insights in the problems being explored. 

By utilizing quantum entanglement distribution, quantum networks are being used to develop secure communication channels for governments, banks, and more, with applications of Quantum Key Distributions (QKDs)~\cite{scarani2009security,nadlinger2022experimental,zhang2024continuous}. Qubits have two properties that enable these advances superposition and entanglement, that allows them to encode and carry information among two entangled qubits. However, qubits can be fragile and can loose their quantum states if there is loss or background noise. Various techniques can measure performance and stability of quantum networks such as state tomography~\cite{cramer2010efficient}, used to monitor the state of all qubits in a network, but is resource-intensive and becomes impractical for larger networks, or process tomography~\cite{mohseni2008quantum}, which involves quantum operation in its fidelity and errors. Researchers have also developed the Bell state monitoring~\cite{zheng2019practical,roos2004bell} method, that utilizes entangled Bell states, where changes in correlations between entangled qubits reveal network errors. This is less resource-intensive than full tomography but offers limited information.

Other developments in quantum repeaters and controllers are being used to amplify weaker signals, increase longer distances or perform qubit operations like rotations, measurements, and entanglement creation.

%\MK{Add a table of current qn testbeds}
Table~\ref{Tab:QNTestBeds} shows some examples of existing quantum network testbed developments. Testbeds offer a real-world platform for experimenting with actual hardware and protocols.

\begin{table*}
\begin{center}
\begin{tabular}{ |p{6cm}|p{8cm}| }
 \hline
 \multicolumn{2}{|c|}{QN Testbeds being Developed} \\
 \hline
 Testbed names & Authors \\
 \hline
 EPB Quantum Network (deployed and in operation) & EPB Chattanooga, Tennessee\\ \hline
 Oak Ridge Quantum Network Testbed & Oak Ridge National Lab, Tennessee\\ \hline
 Center for Quantum Networks (CQN) & Tuscon, Arizona\\ \hline
 Boston- Area Quantum Network (BARQNET)& MIT, Harvard,~\emph{et al.} \\ \hline
 MIT quantum Network testbed & Boston\\ \hline
 Chicago Quantum Exchange (CQE)& Chicago, Illinois\\ \hline
Quantum Application Network testbed for Novel Entanglement Technology (QUANT-NET)    & Lawrence Berkeley National Laboratory, Berkeley, California  \\ \hline
  AFRL Quantum Network &  Rome, Italy\\ \hline
   NYSQIT  &   Stony Brook , BNL,~\emph{et al} \\ \hline
   NICT Quantum Network & National Institute
of Information and Communications Technology\\ \hline
   QuDIT & Lawrence Livermore National Laboratory\\ \hline
   DC-Qnet& Washington DC\\ \hline
   Los Alamos Quantum Network & Los Alamos National Laboratory\\
 \hline
\end{tabular}
\caption{List of quantum testbeds being developed across the world, to name a few.}
\label{Tab:QNTestBeds}
\end{center}
\end{table*}

%QUANT-NET: https://quantnet.lbl.gov/
%Boston-Area Quantum Network (BARQNET): https://arxiv.org/abs/2307.15696
%The New York State Quantum Internet Testbed (NYSQIT): https://www.stonybrook.edu/commcms/CDQP-Inaugural-Workshop/About_the_Center/Testbed.php#:~:text=Testbed%20Infrastructure%3A,quantum%20memories%20and%20entanglement%20sources
%NICT: https://opg.optica.org/jocn/fulltext.cfm?uri=jocn-16-1-A24&id=544187
%QuDIT: https://www.osti.gov/biblio/1834483

\subsection{Application of Quantum Networks}

%\MK{Add table of Q applications in theses area over the last 5 years} % OB
\begin{table*}
\begin{center}
\begin{tabular}{ |p{6cm}|p{8cm}| }
 \hline
 \multicolumn{2}{|c|}{QN Applications over the last 5 years} \\
 \hline
 Application & Description \\
 \hline
   Quantum Key Distribution & Photons used to securely share encryption keys  \\ \hline
   Quantum sensing  &  Measure magnetic fields over large distances with high precision for example~\cite{khalid2023quantum} \\ \hline
   Secure Cloud Computing & Secure access to quantum computers in the cloud\\ \hline
  Distributed Quantum Computation & Distributed quantum processing across geographically distributed quantum computers\\
 \hline
\end{tabular}
\caption{High level examples of Quantum Applications.}
\label{Tab:QNExampleApps}
\end{center}
\end{table*}

Some quantum applications, as seen in Table~\ref{Tab:QNExampleApps}, include Quantum Key Distribution (QKD), sensing or quantum routing. Examples include QKD implementation in NS3 network simulator~\cite{soler2024qkdnetsim+}, clock synchronization accuracy~\cite{spiess2023clock} like GPS~\cite{haldar2023synchronizing} and financial trading~\cite{nande2023quantum}, or measuring magnetic fields or temperature over large distances with high precision~\cite{mullin2023quantum,bao2023quantum}, or quantum-enhanced MRI~\cite{wang2023quantum}. Quantum is also being explored for secure Cloud Computing~\cite{irshad2023iot,sundar2023quantum} applications to benefit fields like materials science and drug discovery~\cite{kumar2024recent,pulipeti2023secure}. 

\section{Overview of Quantum Networks simulators}

%\MK{
%What are the main approaches to simulation, discrete evenet or mathematics driven
%"Most network simulators cited in the literature, such as Omnet++ [8], NetSim [9], and NS-3 [10], are discrete-event simulators"

%Add table of comaprison of simulation toolkits
%Example papers using it...
%}
Omnet++~\cite{varga2010omnet++}, NetSim~\cite{veith1999netsim}, and NS-3~\cite{carneiro2010ns} are well known network discrete-event simulators. These are a series of discrete event modeling toolkits used to simulate traffic flow, customer interactions in a call center, or packet flow in a network. The advantages of these are easier to implement and computationally efficient for large systems. However, these techniques are limited to discrete changes and may not capture continuous behavior accurately.

On the other hand, mathematics-driven simulators or Equation-Based Modeling (EBM) use mathematical equations to represent system dynamics. The advantages of such a simulation approach is that it allows for accurate representation of continuous processes and is flexible, but can be complex to build. For instance, capturing all the possible states and transitions of qubits in the network can be cumbersome and resource intensive. Hamiltonian-based models are another model that describes the energy structure of the network, good for understanding energy transfer, entanglement generation, and other quantum phenomena in the network. Most quantum simulators have been listed in Table \ref{Tab:QNSim}. 

\begin{table*}
\begin{center}
\begin{tabular}{ |p{3.7cm}|p{1cm}|p{2.7cm}|p{2.7cm}|p{4cm}| }
 \hline
 \multicolumn{5}{|c|}{Quantum Network simulators comparison} \\
 \hline
 Simulators name & year & Small topology & Large topology & Focus \\
 \hline
  Squanch~\cite{bartlett2018distributed} & 2018 & & X & GPU Acceleration + Quantum information processing\\
  SeQUeNCe~\cite{wu2021sequence,kettimuthu2019sequence} & 2019 & X & X & Open-Source + User-Friendly Interface \\
  ComNetsEmu~\cite{fitzek2020computing} & 2020 & X & & In-network Artificial Intelligence + hybrid quantum network \\
  QuNetSim~\cite{diadamo2021qunetsim} & 2021 & X &  & Scalability + Performance\\
  NetSquid~\cite{coopmans2021netsquid} & 2021 & X & X & High-Performance Computing Integration \\
  QDNS~\cite{ceylan2021qdns} & 2021 & X &  & Python based + Reusable network protocols \\
  QuISP~\cite{satoh2022quisp} & 2022 & X & X & Error Correction + Fault Tolerance \\
  SimQN~\cite{chen2023simqn} & 2023 &  & X & User-Defined Noise Models \\
  Cisco Qnetlab~\cite{CISCO} & 2024 & X & X & Simulations in the cloud\\
 \hline
\end{tabular}
\caption{Simulation toolkits for Quantum Networks (More details can be seen in Figure~\ref{fig:Networkflowchart}).} %\MK{Update with all simulators in text,m add year of development, Reference to paper of group who developed it.}}
\label{Tab:QNSim}
\end{center}
\end{table*}

%Add diagram of open/commercial quantum network simulators...Flowchart

\begin{figure*}[!htb] 
  \centering
  \includegraphics[width=1\linewidth]{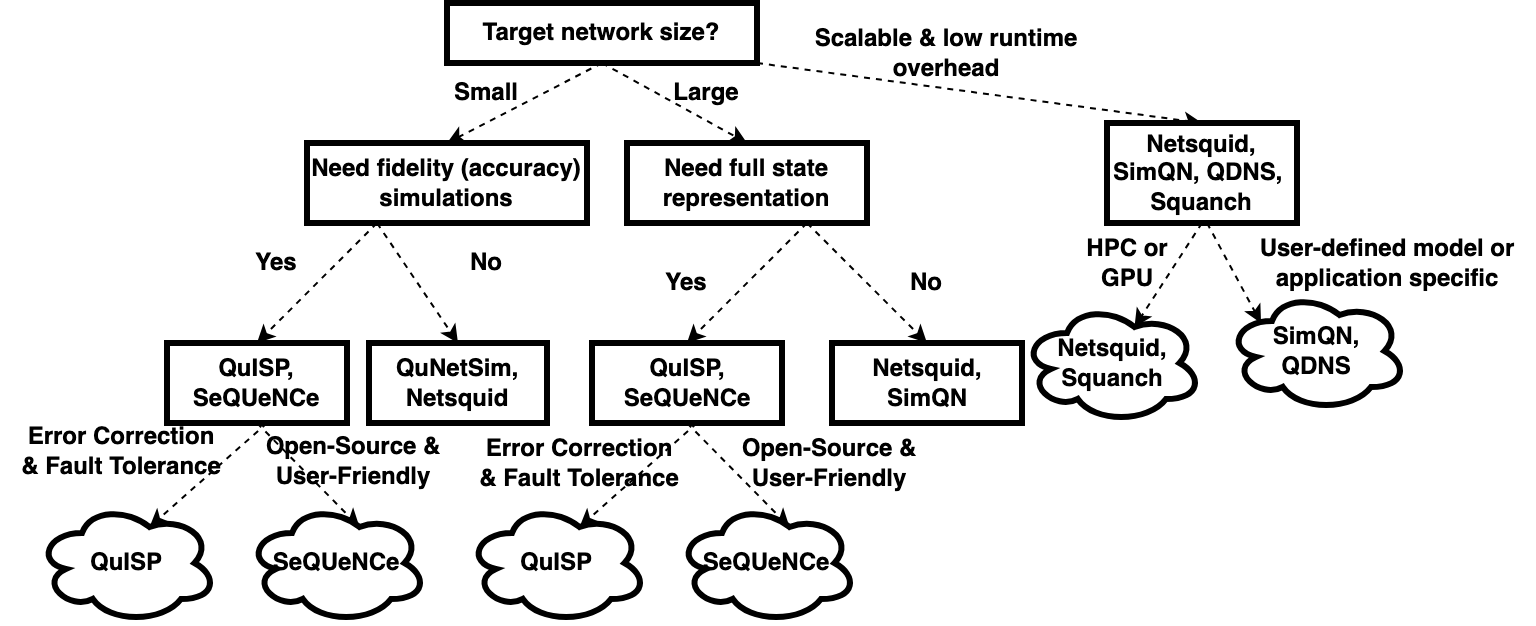}
  \caption{Quantum network simulator flowchart.}
  \label{fig:Networkflowchart}
\end{figure*}

\subsection{QuISP}
QuISP~\cite{satoh2022quisp} is a quantum network simulator that uses the OMNET++~\cite{varga2010omnet++} framework as the base. OMNET++ is a C++ component-based framework used for building network simulators. In 2023, OMNET++ added several models to help users simulate more types of network and routing protocols such as routing protocol for low power and lossy networks. This uses Monte-Carlo simulation and supports error channels including Pauli channels and excitation/relaxation channels. Here, the evolution of the qubit is modeled as a Markov process, encoding flying qubits onto single photons to handle communication between nodes. 

This simulator enables users to experiment with different configurations under realistic and noisy conditions. As such, users can experiment with thousands of qubits in hundreds of nodes, enables more topology types and measure performance and behavior of the network. 
However, the simulation needs to be repeated to gain accurate information on the real state of the qubit and its fidelity. The full-density matrix simulation is not viable for 2G and 3G types of networks. QuISP tackles this issue by assuming that the desired quantum state is known, which is not always the case. 
%https://ieeexplore.ieee.org/abstract/document/9951186
%https://apps.dtic.mil/sti/trecms/pdf/AD1209512.pdf
%https://omnetpp.org/

\subsection{SeQUeNCe}
SeQUeNCe~\cite{wu2021sequence,kettimuthu2019sequence} is a module-based simulator used to simulate new and upcoming quantum technologies. This simulator focuses more on customizability by having five distinct modules that users use to simulate hardware models, entanglement management protocols, resource management, network management, and applications. This structure is similar to the one used by NS3, a network simulator that is frequently used in research for communication networks.

The benefit of this simulator is that we can explore hybrid classical and quantum networks. They can simulate classical channels using quantum network protocols. They also offer the ability to use classical control software to enable rapid experimental progress.

However, the drawback of the simulator is that it assumes there is no loss and perfect reliability for the classical channel always exists. This means that the behavior is uncertain if the network ends up saturated or if there are any failures on the network. Additionally, they assume that the host node is the initiator of any request sent in the application module. Essentially this means that having responses sent back to the initiator of a request may need to have some modifications to be done in the application module of the simulator. 

\subsection{QuNetSim}

QuNetSim~\cite{diadamo2021qunetsim} is a Python-based simulator focused on developing an easy-to-use interface for simulating quantum networks at the network layer. This simulator can also be used to simulate new and existing link layer protocols.

The first benefit of QuNetSim is the ease of use of its interface. It comes with a pre-existing toolbox of quantum protocols that can be implemented in a few lines. This allows for anyone to start learning about quantum protocols and building their own. The second benefit is that it simulates real-time, unlike other simulators. This makes it well-suited to control laboratory hardware.

The drawback of this simulator is similar to the previous two simulators where it assumes links and the physical layer deliver error-free bit and qubit transmission capabilities, which is not the case especially as traffic grows between two points the network should start experiencing more loss and interference.

\subsubsection{Use of Quantum Network simulator as part of Hybrid Network simulator (ComNetsEmu)}

ComNetsEmu~\cite{fitzek2020computing} is a network simulator that is compatible with the QuNetSim simulator. The ComNetsEmu has been built to support in-network Artificial Intelligence (AI). In recent years, an instance of the quantum network simulator QuNetSim has been merged into the link layer of the classical network emulator ComNetsEmu to create an initial hybrid network simulator~\cite{diadamo2021integrating}. The resulting simulator was used as a way to describe the structure of the full stack of a hybrid quantum-classical network.

\subsection{Netsquid}

Netsquid~\cite{coopmans2021netsquid} is a discrete-event-based platform for simulating all aspects of quantum networks. It is widely used across the research community, especially for research in quantum networking and applications. It offers a realistic platform that enables users to collect realistic data on how well their project runs on a quantum network. 

This simulator allows for using large amounts of threads that reduce the runtime linearly with the number of processing cores available. The authors assume that there is sufficient memory available and reiterate that there is a need for a tightly integrated classical control plane with the quantum network. Compared to other simulators, NetSquid also models noise measured during data transfer in the network.

\subsection{SimQN}
SimQN is a discrete-event-based network simulation platform for quantum networks. It is designed to be a functional and easy-to-use tool similar to NS3 for classical networks. It provides researchers with a platform to experiment and simulate quantum networks without needing expensive hardware or waiting for new technology to become available. It aims to address the lack of validation methods for existing quantum network simulations. As explained by Chen~\emph{et al.}~\cite{chen2023simqn}, SimQN provides an average of 8.02 times performance improvement over NetSquid. Another benefit of this simulator is that it provides a modular design that allows for easy customization and extension. This is a benefit since quantum is an evolving field that consistently has new hardware and protocols being developed. As such SimQN enables rapid prototyping and testing of new quantum network protocols and is publicly available~\cite{SimQDNSite}. 

% Title SimQN: A network-layer simulator for the quantum network investigation
%            URL https://ieeexplore.ieee.org/abstract/document/10024900/
%           Year 2023
%      Citations 6
%       Versions 2
%     Cluster ID 4261256352718225778
%       PDF link https://ieeexplore.ieee.org/iel7/65/7593428/10024900.pdf
% Citations list https://scholar.google.com/scholar?cites=4261256352718225778&as_sdt=2005&sciodt=0,5&hl=en
%  Versions list https://scholar.google.com/scholar?cluster=4261256352718225778&hl=en&as_sdt=0,5
%        Excerpt … To this end, we construct our quantum network simulator, SimQN 1, and introduce various innovative designs in its architecture. The first design is the multiple fine-grained physical …

% More simulators that we should look at:
% I used the query code located at https://github.com/fjxmlzn/scholar/blob/master/scholar.py to query google scholar

\subsection{QDNS}

QDNS~\cite{ceylan2021qdns},  Quantum Dynamic Network Simulator, is a theoretical framework for simulating the behavior of qubits. Similar to SimQN, QDNS is publically available on~\cite{QDNSSite}. It is a Python-based simulator which makes it user-friendly but reduces the resulting performance of the simulator compared to other simulators. It also provides reusable network protocols and utilities which reduces development time and effort for researchers.

%          Title QDNS: Quantum Dynamic Network Simulator Based on Event Driving
%            URL https://ieeexplore.ieee.org/abstract/document/9654307/
%           Year 2021
%      Citations 0
%       Versions 0
%       PDF link https://ieeexplore.ieee.org/iel7/9654277/9654279/09654307.pdf
%        Excerpt … These events were the main reason to encourage us to develop a quantum network simulator. It was also an additional motivate for us to introduce new functions that we thought were …
\subsection{Cisco's Qnetlab}
This is a software development kit created by Cisco Research~\cite{CISCO} to simplify the design and execution of simulations for quantum networks. It provides a platform for building and running simulations in the cloud, as well as a protocol builder for designing quantum network protocols.

\subsection{Squanch}
Squanch stands for Simulator for Quantum Networks and CHannels~\cite{bartlett2018distributed}. This is an open-source Python library for simulating quantum networks, focusing on providing tools for developers to design and run simulations, not on building actual quantum networks.

Squanch is specifically designed for simulating distributed quantum information processing, making it particularly efficient for modeling complex quantum networks that involve multiple nodes and communication channels. By running simulations with different parameters, researchers can fine-tune existing protocols to improve their performance and reliability. Additionally, Squanch can be used to explore the potential of novel network designs and assess their feasibility before committing resources to building them in hardware. Finally, the library leverages the power of multiple processors to run simulations faster, especially for large-scale networks.

Unlike real-world quantum systems, Squanch simulations are based on models and approximations. These models might not capture all the complexities of real-world systems, leading to potential inaccuracies in the simulation results. This is especially true for simulating large-scale networks or complex noise models. While simulations can provide valuable insights, there's still a gap between simulated behavior and the actual performance of quantum hardware. Factors like hardware limitations, control errors, and decoherence can significantly affect the practical implementation of protocols designed based solely on simulations. Finally, although Squanch utilizes parallelization, simulating extremely large or intricate networks can still become computationally expensive and time-consuming. This can limit its applicability for studying certain types of large-scale quantum systems.

\section{Background and Discussions}
%\MK{Why simulation of QN is good}
%You can add repeaters modeling is importnat due to distance lack...
Building and experimenting with quantum networks can be expensive and time-consuming. Simulations allow researchers to test and refine their designs virtually, exploring different architectures, protocols, and parameters to optimize network performance before committing to physical implementations, crucial for developing robust and reliable quantum communication and computing systems.

\subsection{Model Quantum Devices}
Quantum devices are devices that exploit the principles of quantum mechanics to perform computations. Examples include superconducting qubits~\cite{devoret2004superconducting,kjaergaard2020superconducting}, trapped ion qubits~\cite{bruzewicz2019trapped}, and photonic qubits~\cite{o2009photonic,nisbet2013photonic,liu2022demand} used in quantum computers. Superconducting qubits are used in superconducting circuits cooled to near absolute zero to encode quantum information. Trapped ion qubits are used individual ions held in place by electromagnetic fields to represent quantum information. Photonic Qubits use photons (particles of light) to encode quantum information.

\subsection{Model Quantum Protocols}
There are various techniques to model quantum protocols including Stochastic Simulation~\cite{apolloni1989quantum,yang2018matrix,wu2023implementing} and Tensor Network Methods~\cite{huang2023tensor,banuls2023tensor}. Stochastic simulation employs a random sampling approach to simulate the evolution of the quantum state during a protocol. This technique is efficient for analyzing protocols with limited resources. However, it can be statistically noisy such that this approach might not capture the full complexity of the system. Another commonly used approach is the Tensor Network Method, which is a technique that represents the quantum state by decomposing it into a network of tensors. This approach offers an efficient way to model systems with entangled qubits, especially for specific types of entanglement structures. These techniques can become computationally expensive for more general scenarios.

\subsection{Model Quantum Repeaters}
When developing repeaters, it is important to understand that the type of links used would impact the qubit generation technique used by the network. Researchers also need to understand when the resource allocation policy should be triggered and what factors seem to be crucial in the development of a quantum network.

First, the placement of repeaters has a direct impact on the loss, similar to the placement of the GnB nodes (relay antennas) in a wireless or 5G network. If the relay antennas are too far apart, then the signal may suffer from more interference since it needs to travel farther. In quantum networks, information degrades over distance due to decoherence. This means that the distance between repeaters should be less than the decoherence length of the chosen channel (fiber optic cable, free space, etc). 

Additionally, not all repeaters are built the same way. The quantum memory type and lifetime vary between repeater technologies. This means that some repeaters can transmit signals across a wider distance than others. The entanglement generation and manipulation capabilities and the maintenance also vary between different repeater technologies. A network architect needs to understand the capabilities of each type of repeater used in the network before determining where to place them. This is where simulators can become useful since they give the user the ability to quickly try different repeater technologies. They can then quantify each topology and select the best one for the network's purpose.

% - How can one model the placement of repeaters?
One approach to model repeaters is using a Heuristic or Analytical model. These algorithms use simplifying assumptions to quickly find possible repeater placements that can satisfy minimum requirements like connectivity and cost. One can expand on this to incorporate detailed channel characteristics, technology specifications, and network topology constraints to optimize placement for factors like fidelity, throughput, and cost. However, while those models can be easy to implement, they struggle when modeling complex systems. To tackle these issues there are simulation tools in which quantum repeater networks can be simulated to evaluate different placement scenarios and their performance under various conditions.

\subsection{Complexity of Modeling Quantum Networks}
%levels of abstraction
Simulating entire quantum networks at the lower levels of individual qubits quickly becomes computationally intractable as the network grows~\cite{pan2022simulation}.  Abstraction levels allow researchers to focus on the essential features of the network for a specific task~\cite{wehner2018quantum}. Additionally, abstraction levels provide a framework for designing and developing quantum protocols and algorithms without worrying about the intricate workings of the underlying physical qubits. 

%complexity
For example, modelling a long-distance quantum network can reveal the error rate (complexity~\cite{siomau2016structural,hayashi2007quantum}) increases rapidly with distance. Adding a sophisticated error correction protocol or exploring alternative network architectures with built-in redundancy for enhanced resilience can help develop better models. Complexity models can be used to simulate how errors propagate through the network and assess the effectiveness of different error correction protocols. The resulting complexity analysis helps design more robust networks with improved fidelity (accuracy) in transmitting quantum information.

%\subsection{Performance of Quantum Simulators}

\textit{Compuational Resources Needed.} Simulating the dynamic behavior of quantum networks, where nodes can be added or removed, and connections can change over time, is also of interest. This includes nodes that become unresponsive because of some fault, allowing quantum network simulators to become more realistic. Understanding how the performance of the network topology changes when a node is unavailable can help researchers understand how their protocol and network topology work in the real world.

Optimizing the use of quantum resources, such as entanglement generation and distribution, is also a key challenge. At the core of quantum networks, entanglement generation and distribution are used to transmit data from one node to another. Depending on the interaction between qubits and the outside world the resulting qubit value by a node may change and understanding how that change happens would lead simulators to become more realistic tools for researchers.

\textit{Benchmarking Simulations.}
These include entanglement generation success rates~\cite{vinay2019statistical, patil2022entanglement}, fidelity of entanglement~\cite{li2022fidelity, guhne2021geometry}, entanglement distribution speed~\cite{ikuta2018four}, or  scalability (ex: running millions of agents at the same time with massive data at HPC speed). Other metrics involve the need to integrate with MPI to run on HPC system to increase parallelism and avoid bottlenecks. With these available benchmarks, there is a need to create a tool that can give an overall view of the performance of their network topology and protocols.

Often, benchmarks have focused on measuring fidelity and the resource usage. Fidelity is a measure of how well the generated entangled states match the ideal entangled states and the benchmarks assess the fidelity of entanglement distribution, and the quality of the generated entangled pairs. Additionally, evaluating the time required to establish entanglement between nodes also helps assess the efficiency of entanglement generation processes.

\section{Other Approaches to Qubit Behavior Simulation}

There are several methods used to simulate qubits such as Stochastic Wavefunction~\cite{breuer1999stochastic}, Tensor Network~\cite{montangero2018introduction}, and Monte-Carlo~\cite{hammersley2013monte}. Stochastic wavefunction methods, like the Quantum Trajectory Approach, involve random sampling of quantum trajectories to approximate the evolution of a quantum system. These methods can be more computationally efficient than exact simulations for certain scenarios. Tensor network methods, including the Matrix Product State~\cite{perez2006matrix} (MPS) and the Tensor Network State~\cite{evenbly2011tensor} (TNS) representations, are used for simulating the state of a quantum system. These methods offer a more efficient representation of quantum states, especially for systems with entanglement. And, Monte-Carlo methods are employed in quantum Monte-Carlo simulations to estimate physical quantities related to quantum systems. Variational Monte-Carlo and Diffusion Monte-Carlo are examples of techniques used for qubit simulations. Since currently there is still development happening on different generation methods for qubits, quantum network simulators should at least offer all three approaches so that researchers can explore the effect of all three methods on their new protocol. 

\subsection{Understanding Qubit behavior}
In this section, we focus on developing an analytical model that can be used to model the evolution of the density matrix of a quantum system, such as a qubit, going through a quantum communication channel. We start with qubit transmission since it is the base usage of networks and qubits are quantum systems that carry the quantum information from one node to another in a quantum network. A single qubit state can be expressed by using the Dirac notation~\cite{tumulka2009dirac,hidary2021dirac}. An example representation of a qubit is represented in Equation~\ref{qubit}.

%The first approach to modeling the expected outcome of a quantum simulator is to use an analytical model. In this section, we will focus on building an analytical model that can be used to model the evolution of the density matrix of a quantum system such as a qubit going through a quantum communication channel. We start with this part of the network since the qubit is what will carry the quantum information from one node to another in a quantum network. A single qubit state can be expressed by using the Dirac notation~\cite{tumulka2009dirac,hidary2021dirac}. An example representation of a qubit is represented in Equation~\ref{qubit}.
\newcommand{\ket}[1]{\left \rvert #1 \right \rangle}

\begin{equation}\label{qubit}
    \ket{\Psi} = \alpha \ket{0} + \beta \ket{1}
\end{equation}

Researchers have used the the Bloch equation~\cite{hammes2023behaviour,wie2020two,ruskov2003spectrum}, as described in Equation~\ref{Bloch}, to model qubit behavior. This equation describes the evolution of the Bloch vector, which is a three-dimensional vector that represents the state of a qubit. It uses differential equations that takes into account the Hamiltonian of the qubit system, which is the operator that describes the energy of the system, which includes 2 or more qubits. It can also be used to describe other aspects linked to qubit behaviors such as precession and relaxation. Precession is the rotation of the qubit state around the Bloch sphere at a frequency that is proportional to the qubit transition frequencies. Relaxation is the decay of the qubit state to its equilibrium value over time due to interactions with the environment. 

%\begin{equation}\label{qubit}
\begin{equation}\label{Bloch}
\begin{split}
dMx/dt = \omega{yMy} - \omega{zMz}\\
dMy/dt = -\omega{xMx} + \omega{zMz}\\
dMz/dt = -\omega{xMx} - \omega{yMy} - (Mz - M0)/T1
\end{split}
\end{equation}

$Mx, My,$ and $Mz$ are the components of the Bloch vector on the Bloch sphere. $\omega$x, $\omega$y, and $\omega$z are the frequencies of the qubit transitions in the $x, y,$ and $z$ directions, respectively. M0 is the equilibrium magnetization, which is typically zero for qubits. T1 is the longitudinal relaxation time constant, which is the time it takes for the longitudinal magnetization to relax back to its equilibrium value after it has been perturbed. 

Overall, Bloch equations have been used to model qubit behaviors on quantum communication channels~\cite{lang2017topological,cacciapuoti2020entanglement,ban2005decoherence}. However, they are not a fully accurate representation of qubit dynamics~\cite{li2004quantum,perego2020coherent}. This is due to their classical nature, limited scope to two-level systems~\cite{skinner2003application,li2017exact}, and lack of consideration for other quantum aspects such as decoherence~\cite{schlosshauer2019quantum}. For more accurate modeling, more sophisticated techniques such as the master equation are typically used. Therefore, the master equation can be used to determine the evolution of the density matrix of a quantum system.

The master equation for a qubit in a quantum communication channel is given by Equation~\ref{Master}, where $\rho$ is the density matrix of the qubit, L is the Hamiltonian of the qubit system, $\Gamma$ is the decoherence rate, and $\rho{eq}$ is the equilibrium density matrix of the qubit. By using this equation, we can directly model the impact of different types of channel parameters on the resulting density matrix of the qubit. $\Gamma$ and L can be used to experiment with different rates of noise on a quantum system.

\begin{equation}\label{Master}
    d\rho/dt = -iL\rho - \Gamma(\rho - \rho{eq})
\end{equation}

The decoherence rate value $\Gamma$ depends on the type of noise that the qubit is subject to and the properties of the environment. The decoherence rate can be calculated from the longitudinal relaxation time constant (T1) and the transverse relaxation time constant (T2), as described by Equation~\ref{decoherence}. By looking at this equation and the Bloch equation we can see that the Bloch equation is missing the decoherence value when calculating the updated state of the qubit.

\begin{equation}\label{decoherence}
    \Gamma = 1 / (2 * T1) + 1 / (T2 * \pi)
\end{equation}

The Hamiltonian is written as a formula showing the interaction between the energy of 2 qubits described as $L = L_0 + L_{int}$. $L_0$ is the free Hamiltonian, which describes the energy of the qubits in the absence of any interactions. On the other hand, $L_{int}$ is the interaction Hamiltonian, which describes the energy of the interactions between the qubits. Both $L_0$ and $L_{int}$ are described in Equation~\ref{Free and int}.

\begin{equation}\label{Free and int}
\begin{split}
L_0 = (1/2) * \Delta\omega * \sigma_z\\
L_{int} = J * \sigma_i \sigma_j
\end{split}
\end{equation}

In the equation, the variables include different components of qubits. $\Delta\omega$ is the qubit transition frequency. $\sigma_z$ is the Pauli Z operator~\cite{setia2020reducing,heidari2023learning}. J is the coupling constant between the qubits~\cite{wang2010classical,piela2013ideas}. There exist multiple models that can be used to determine the J coupling value, such as Ising Model or XY Model. It depends on the type of interaction that is being modeled. For example, the XY model seems to depend heavily on the temperature of the system while the Ising model the intrinsic dimension is independent of the real-space topology~\cite{vitale2023topological}. Therefore, an interface that uses the equation set needs to have the ability to swap the coupling constant depending on the need of the user and the experiment. Finally, $\sigma_i$ and $\sigma_j$ are the Pauli operators for the ith and jth qubits, respectively. The Pauli operators are three operators: the X, Y and Z operators.  The Pauli operators~\cite{djordjevic2021quantum} are denoted using the formulas described in equation~\ref{Pauli}. Because, even the smallest components of these equations can be written as vector math, using the existing literature, we can build a model that can be used by other researchers to validate new development in quantum network simulators.

\begin{equation} \label{Pauli}
  \begin{split}
    X = |0><1| + |1><0| = 
    \begin{bmatrix}
    0 & 1 \\
    1 & 0 
    \end{bmatrix} \\
  Y = -i |0><1| + i |1><0| =
    \begin{bmatrix}
    0 & -i \\
    i & 0 
    \end{bmatrix} \\
  Z = |0><0| - |1><1| =
    \begin{bmatrix}
    1 & 0 \\
    0 & -1 
    \end{bmatrix} \\
\end{split}
\end{equation}

Now that the behavior of a qubit is shown using vector math, it is important to acknowledge that such an approach does not scale well. The number of variables and equations involved in analytical models grows exponentially with the number of qubits or network components. This can quickly render them impractical for simulating realistic quantum networks. This means that validating a simulation of a full network may need other validation techniques such as using Monte-Carlo simulations, which enable us to run the simulation repeatedly under different conditions. Because of the scaling problem, translating analytical models into efficient and accurate simulation code can also be a complex task, which can lead to the potential introduction of errors or numerical instabilities.

Additionally, assumptions taken by the developers of the analytical models play a crucial part in the outcome of the validation. Verifying these assumptions in real-world quantum networks can be challenging, leading to uncertainties in the model's accuracy. Additionally, analytical models often make simplifying assumptions, such as perfect qubits, noiseless operations, and ideal channels. These assumptions may not accurately reflect the behavior of real-world quantum networks, which are inherently susceptible to noise, imperfections, and environmental influences. As a result, analytical models may fail to capture the full complexity and nuances of physical quantum systems, potentially leading to inaccurate or misleading simulation results.

\subsection{Analytical Benchmarking}
Dealing with multiple dimensional vector data, a first approach for validation is Euclidean distance~\cite{krislock2012euclidean}. This metric lets users visualize how closely the simulated qubit density matrix matches the expected one. A small distance indicates that the simulated channel behaves as expected in a real quantum network running the same workload.

However, for a deeper understanding of the discrepancies between the two density matrices, the Earth Mover's algorithm, can be used for comparing 2D data, needed to transform one distribution into the other. By combining Euclidean distance and the Earth Mover's value, users can gain a comprehensive picture of how well the simulation replicates reality. A small distance and a low `dirt' value would strongly validate the resulting qubit value.

Besides Euclidean distance and Earth Mover's distance, there are several other comparison metrics for 3D vector data. These comparison metrics include distance-based metrics, distribution-based metrics, and directional metrics. We want to use a combination of all three types of metrics to validate the simulated qubit behavior. Similar to Euclidean, the Manhattan distance allows distance metric calculated over the sum of absolute differences instead of squares, emphasizing outliers. Outliers are important to identifying any isolated discrepancies in the simulated results. Using the Root Mean Square Distance between the simulated qubit density metric and the theoretical density metric, it records the impact of larger errors.

Another benchmark is to understand how the simulation modeled the distribution of the density matrix. The Kullback-Leibler divergence~\cite{kullback1951kullback} (KL divergence) measures how much information is lost when describing one distribution with another. Jensen-Shannon divergence~\cite{menendez1997jensen}, is another distribution-based metric, similar to KL divergence but symmetric based, useful for comparing two distributions simultaneously. This metric allows users to evaluate how well the simulation models the interactions between 2 qubits. Also one can combine the result of Jensen-Shannon divergence with Bhattacharya distance~\cite{kailath1967divergence,choi2003feature} to see how much overlap the simulated matrix has with the resulting theoretical matrix. The Bhattacharya distance measures the overlap between two probability distributions. 

Finally, comparing the direction of both the simulated density matrix and the resulting density matrix one can compare the angle between both vectors. This measures the rotation needed to align one vector with the other. The `size' of the rotation needs to align one set of vectors with the other, we can use the Rotation Matrix Frobenius norm.

\section{Conclusion and Future Directions of Research}
This survey identifies a need for a standardized platform to validate existing quantum simulators. By using existing theoretical work for a validation interface, one can generate values of the density matrix that would be expected depending on certain network conditions. Developing a user-friendly interface that can be used by the research community, can help users validate the result of simulators. 

Further, this can also be used to study the interaction between classical and quantum networks. Because qubits carry data in quantum networks it is important to understand how the interactions impact the qubit values. The validation interface should evolve to include the validation techniques for the behavior of hybrid networks expected in the real world.

We also discussed using analytical models such as modeling a qubit behavior, quantum bnchmarking, or classical shadow optimization to verify quantum network simulators. For quantum benchmarking, we propose to build a library of benchmarking workloads and protocols that can be used to probe the capabilities and limitations of quantum devices and simulators. These benchmarks, like Bell state preparation and tomography, provide a reference point for evaluating the accuracy and fidelity of simulations. For the case of classical shadow optimization, we can use a classical optimization algorithm that mimics the behavior of a quantum algorithm. This algorithm can then provide a classical approximation of the quantum simulation results. Comparing the two sets of results can highlight potential errors or limitations in the simulator. 

Further explorations of formal verification testing and code review can be performed, with sensitivity analysis, to include noise injection and statistical analysis to understand how the simulation is performing. Simulating the effects of different types of noise, such as decoherence and state preparation errors, is crucial for assessing the robustness and accuracy of quantum network simulations. It can help researchers identify sensitive areas and potential sources of error in the simulation. Additionally, performing a parameter sweep by systematically varying key parameters of the quantum network, such as coupling strengths or interaction times, allows researchers to study the behavior of the simulated system across a wider range of conditions. This can reveal unexpected dependencies or edge cases that require further validation.

Finally, using existing validation techniques such as Monte-Carlo simulations can help reveal inconsistencies in the outcome of the simulation. Running the simulation multiple times with different initial conditions or random noise variations helps to quantify the statistical uncertainty and variability of the results. This provides a more complete picture of the simulated system's behavior and avoids concluding single realizations. Additionally, statistical tests can be used to formally compare the simulated outcomes with a particular hypothesis or reference data. This allows for a rigorous assessment of the validity and significance of the simulation results.

\section{Acknowledgements}

During the preparation of this work the author(s) used Google Bard in order to improve grammar and flow of the text in the paper, and to supplement background research. After using this tool, the author(s) reviewed and edited the content as needed and take(s) full responsibility for the content of the publication. This work was supported by the U.S. DOE Office of Science, Office of Advanced Scientific Computing Research, under award 66150: ``CENATE - Center for Advanced Architecture Evaluation'' project and the DOE  ASCR Early Career Grant ``Large Scale Deep Learning for Intelligent Networks'' award ERKJ435 hosted at Oak Ridge National Laboratory. The Pacific Northwest National Laboratory is operated by Battelle for the U.S. Department of Energy under contract DE-AC05-76RL01830. 

This manuscript has been authored by UT-Battelle, LLC, under contract DE-AC05-00OR22725 with the US Department of Energy (DOE). The US government retains and the publisher, by accepting the article for publication, acknowledges that the US government retains a nonexclusive, paid-up, irrevocable, worldwide license to publish or reproduce the published form of this manuscript, or allow others to do so, for US government purposes. DOE will provide public access to these results of federally sponsored research in accordance with the DOE Public Access Plan (https://www.energy.gov/downloads/doe-public-access-plan).

\bibliographystyle{unsrt}  
%\bibliography{references}  %%% Remove comment to use the external .bib file (using bibtex).
%%% and comment out the ``thebibliography'' section.

%%% Comment out this section when you \bibliography{references} is enabled.
\bibliography{references}

\end{document}